\documentclass[english]{article}
\usepackage[T1]{fontenc}
\usepackage[latin9]{inputenc}
\usepackage{geometry}
\usepackage{amsthm}
\usepackage{amsmath}
\usepackage{babel}
\begin{document}

\title{Matrix Factorization Method for Decentralized Recommender Systems}

\author{Zheng Wenjie}
\maketitle
\begin{abstract}
Decentralized recommender system does not rely on the central service
provider, and the users can keep the ownership of their ratings. This
article brings the theoretically well-studied matrix factorization
method into the decentralized recommender system, where the formerly
prevalent algorithms are heuristic and hence lack of theoretical guarantee.
Our preliminary simulation results show that this method is promising.
\end{abstract}

\section{Introduction}

Recommender Systems (RS) \cite{ricci2011introduction} are a kind
of system that seek to recommend to users what they are likely interested
in . Unlike search engines, the users do not need to type any keyword.
The RS's will learn their interest automatically. For instance, if
the user has just bought a numeric camera, the RS will recommend to
him some SD memory cards; if a user watches a lot of action movies,
the RS may suggest some other action movies to him. And this is the
typical behaviors which we observe universally in Netflix (movies),
Youtube (videos), Google Play (apps), Facebook (friends), Amazon (goods)
and other platforms today.

Recommender systems play an important role in our daily life. They
target user tastes and profiles and provide them with relevant information.
This keeps users from drowning in the ocean of data, and helps them
quickly find what they exactly want or like. This capacity of providing
users with targeted information is achieved by collecting user-related
data into a central database, and then running various recommendation
algorithms on it.

Here comes the issue. The data, e.g. the ratings given to a movie
by users, are generated by the users themselves. Logically, they are
property of the users who generated them. But why in the end, the
companies which provide the recommender system take possession of
it? What if they sell them to a third party? What if the user privacy
is leaked? A related example is the \emph{Netflix Prize}, the Netflix
company published 100~480~507 ratings that 480~189 users gave to
17~770 movies on the Internet in an anonymous way \cite{bennett2007netflix}.
However, one year later, two researchers from the University of Texas,
de-anonymized some of the Netflix data by matching the data set with
film ratings on the Internet Movie Database \cite{narayanan2008robust}.

Therefore, it would be interesting if we could design a computational
framework where the users can get recommendation without the help
of a recommender service provider, and the users do not even transfer
their data to others. Instead, they keep the data on their own computers
and do all the computation collaboratively with other users in a \emph{decentralized}
way.

In fact, there are already some works in this direction \cite{han2004scalable,kermarrec2010application,massa2007trust,o2005trust,ziegler2005towards}.
They use trust or some similarity metric to build a graph among the
users, and then use some propagation algorithms to do the recommendation.
These methods are straightforward and heuristic. However, they lack
theoretical guarantee. Moreover, recent research has developed many
powerful methods such as matrix factorization \cite{koren2009matrix},
which are incompatible to the network propagation framework.

The goal of this article is to introduce the matrix factorization
method into the decentralized recommender system. Before our effort,
there are already some trials in \emph{Network Distance Prediction}
\cite{liao2010network,liao2013dmfsgd}. This article will focus on
the specialty of recommender systems. We first introduce this model
and the associated theoretical analysis in Section~\ref{sec:model}.
Then, we proceed in Section~\ref{sec:algorithm} to elaborate a decentralized
algorithm to solve the model. Section~\ref{sec:simulation} presents
some preliminary simulation. We end this article with a discussion
in Section~\ref{sec:discussion}.

\section{Matrix factorization model and consistency \label{sec:model}}

In this section, we first present the matrix factorization model of
the RS problem. Then, we show that its estimator is consistent.

Suppose there are $m$ users and $n$ items in the universe. $\Theta_{m\times n}$
is the unknown matrix of true ratings that each user will give to
each item. This is a dense matrix without any missing values. However,
since there are so many items, users are not able to test every item
and their ratings are corrupted by noise. What we finally observe
is a sparse matrix $X_{m\times n}$, which could be regarded as some
approximation of $\Theta$. The objective of RS is to recover $\Theta$
from $X$, so that it can recommend to users the items with the largest
values in $\Theta$.

In order for that, \cite{koren2009matrix} uses a factorization method.
It regards $\Theta$ as some low-rank matrix, i.e. $\textrm{rank}(\Theta)=r\ll\min(m,n)$.
Thus, $\Theta$ can be approximated by the product of two low dimensional
matrices $U_{r\times m}$ and $V_{r\times n}$: $\Theta=U^{T}V$.
This representation has a nice interpretation. The $i$-th column
of $U$ can be seen as the profile of the $i$-th user, and the $j$-th
column of $V$ can be seen as the profile of the $j$-th item; the
rating $\Theta_{ij}$ that the $i$-th user gives to the $j$-th item
is just the inner product of these two profiles.

Then, the estimation of $\Theta$ is equivalent to the estimation
of $U$ and $V$. In this way, \cite{koren2009matrix} proposes the
following method:
\begin{equation}
\min_{U,V}\left\Vert \mathcal{P}_{\Omega_{X}}(U^{T}V-X)\right\Vert _{F}^{2}+\lambda(\left\Vert U\right\Vert _{F}^{2}+\left\Vert V\right\Vert _{F}^{2}),\label{eq:factorization}
\end{equation}
where $\left\Vert \cdot\right\Vert _{F}^{2}$ represents the Frobenius
norm, $\Omega_{X}$ is the support of $X$ and $\mathcal{P}$ is the
projection operator. Besides the obvious approximation term, (\ref{eq:factorization})
includes a regularization term as well.

\cite{recht2010guaranteed} shows that (\ref{eq:factorization}) is
equivalent to the following optimization problem:
\begin{equation}
\begin{aligned}\min_{M} & \qquad\left\Vert M\right\Vert _{*}\\
\text{s.t.} & \quad\left\Vert \mathcal{P}_{\Omega_{X}}(M-X)\right\Vert _{F}\le\rho,
\end{aligned}
\label{eq:nuclear}
\end{equation}
for some $\rho(\lambda)$ depending on the value of $\lambda$, where
$\left\Vert \cdot\right\Vert _{*}$ represents the nuclear norm (a.k.a.
trace norm). And the minimal solution $\widehat{M}$ of (\ref{eq:nuclear})
is the product of the minimal solution $\widehat{U}$ and $\widehat{V}$
of (\ref{eq:factorization}): $\widehat{M}=\widehat{U}^{T}\widehat{V}$.

\cite{fazel2008compressed} proves that if the projection operator
$\mathcal{P}_{\Omega_{X}}$ satisfies \emph{restricted isometry property
}(RIP): $\left(1-\alpha\right)\left\Vert A\right\Vert _{F}^{2}\leq\tfrac{1}{p}\left\Vert \mathcal{P}_{\Omega_{X}}\left(A\right)\right\Vert _{F}^{2}\leq\left(1+\alpha\right)\left\Vert A\right\Vert _{F}^{2}$,
for any matrix $A$ with sufficiently small rank and $\alpha\in\left(0,1\right)$
sufficiently small, where $p$ is the proportion of non-missing values
of $X$, then
\[
\left\Vert \widehat{U}^{T}\widehat{V}-\Theta\right\Vert _{F}\leq C_{0}p^{-1/2}\rho.
\]
This means that the matrix factorization algorithm approximately recovers
the true ratings $\Theta$ with a few corrupted ratings $X$.

Therefore, as soon as a decentralized algorithm converges to the minimal
solution of (\ref{eq:factorization}) (or approximately), it enjoys
a nice theoretical guarantee.

\section{Decentralized matrix factorization algorithm \label{sec:algorithm}}

In this section, we present our decentralized matrix factorization
algorithm. We first clarify the computation environment. Then, we
proceed to describe the algorithm. Then, we explain the relation between
this algorithm and optimization problem (\ref{eq:factorization}).
Finally, we show how to use the trained model to get recommendation.

At the beginning, each user $i$ holds his rating vector $x_{i}$
in his computer. And he also spares some place in the memory to store
the user profile $u_{i}$, which is randomly initialized. To store
the item profile $v_{j}$, we suppose that there are many spared computational
entities scattered on the Internet. Such entity can be routers or
volunteers' PC. For ease of description, we will just use ``router''
to refer to it. The item profile is also randomly initialized. 

Suppose that the user is in some P2P network. He does not know the
global topology of the network, but he possesses a short list of routers
that he can send messages to. This list can be dynamically updated.
It is the same for the routers. They have a short list of users that
they can send messages to. An additional thing is that each user/router
should choose a learning rate $\eta$ for itself. This $\eta$ can
vary between different users/routers. With a bit abuse of notation,
we do not put any super/subscripts on $\eta$. Higher $\eta$ values
stand for that the profiles are updated more aggressively.

Once a user/router is ready, it joins in the network and begins the
communication with other routers/users. Each user/router will basically
do two things: broadcast its profile parameter according to its list,
and receive the broadcast profile parameter to update its own profile
parameter. While for the broadcasting, users and routers do exactly
the same thing, their behaviors are slightly different during the
updating procedure. When user $i$ receives an item profile, say $v_{j}$,
if rating $x_{ij}$ is not missing (i.e. user $i$ does rate item
$j$), then user $i$ updates his user profile:
\begin{equation}
u_{i}\longleftarrow u_{i}-\eta v_{j}(u_{i}^{T}v_{j}-x_{ij})-\eta\lambda u_{i}.\label{eq:uupdate}
\end{equation}
When item $k$ receives a user profile $u_{l}$, if it does not know
the rating $x_{lk}$, then it requests it from user $l$. Then it
updates its item profile:
\begin{equation}
v_{k}\longleftarrow v_{k}-\eta u_{l}(u_{l}^{T}v_{k}-x_{lk})-\eta\lambda v_{k}.\label{eq:vupdate}
\end{equation}

Each user/router can freely join in or quit the learning process at
any time. Therefore, the algorithm is robust against the breakdown
of individual nodes. And when a user gives a new rating to an item
(e.g. he just saw the movie yesterday), the learning process does
not need to restart from scratch; it follows exactly the same protocol.
And it is the same when new users or new items come into the system.
The computation network can keep going on and it scales up well.

To answer the question why this algorithm \emph{can} work well, it
is sufficient to notice that (\ref{eq:uupdate}) and (\ref{eq:vupdate})
is a variant of \emph{stochastic gradient descent} (SGD) algorithm
to minimize the optimization problem (\ref{eq:factorization}). What
should be accentuate is that our algorithm is not exactly SGD. For
a true SGD, (\ref{eq:uupdate}) and (\ref{eq:vupdate}) should be
done at the same moment and at the same place for a pair of profile
parameters $\left(u_{i},v_{j}\right)$. Therefore, its update sequence
can only be something like $\ldots\left(u_{i},v_{j}\right),\left(u_{k},v_{l}\right)\ldots$.
However, in our algorithm, each time, only one profile gets updated,
and a user/router only updates his own $u_{i}$/$v_{j}$. As a consequence,
the evolution of the profile parameters in our algorithm can have
patterns like $\ldots u_{i},u_{k},v_{j},v_{l}\ldots$ , which will
never appear in the true SGD.

The above is the learning process. For the recommendation, say if
user $i$ would like to know whether he will appreciate item $j$,
he just need to request the profile parameter $v_{j}$ from the router
which hosts it. He calculates its inner product with $u_{i}$, and
then he knows what rating he is likely to give to this item. In contrast
to traditional recommender systems, which only give recommendation,
our system actually gives the predicted rating. User can thus get
a rough idea about the item that he is hesitating to purchase or the
movie he is hesitating to watch.

\section{Simulation \label{sec:simulation}}

In this section, we test our algorithm on both synthetic dataset and
real dataset. We will first show the experiment protocol. Then, we
present the database and the result our algorithm yields.

Since all this thing is currently just a concept, we have not yet
a genuine decentralized recommender system that we can experiment
with. We will simulate it in a reasonable way. For this reason, we
make a hypothesis to restrict the behavior that is allowed to happen
in a genuine decentralized network, so that our simulation reflects
more or less the phenomenon in practice. 

We suppose that given two messages $D_{1}$ and $D_{2}$ sent by the
same emitter, where $D_{1}$ is emitted before $D_{2}$, if both are
ever received by a receptor, then the reception of $D_{1}$ always
happens before the reception of $D_{2}$. This assures that a delayed
message sent millions years ago will not erase the effect of a recent
message.

This hypothesis is realistic. We can make sure it happens by timestamping
the data and discarding the delayed data. For each message emitted,
the emitter annotates it in adding the time information and the emitter's
MAC address. If a receiver receives a message with an earlier timestamp
than the last message received from the same emitter, then it simply
discards it.

In this way, we can simulate our algorithm in a single computer in
an efficient way. Indeed, in this case, it is equivalent to repeatedly
sampling a rating $x_{ij}$ and updating either $u_{i}$ or $v_{j}$
(only one of them, not both).

For synthetic data, we let $\Theta$ be a low-rank matrix with floating-point
values close to $\left\{ 1,2,3,4,5\right\} $, and $\textrm{rank}(\Theta)=r=10$.
$X$ is generated by rounding $\Theta$ to the closest integers, and
randomly deleting 20\% values as missing ratings. We test for matrix
size $m=n=100$, $200$ and $500$. The evaluation criterion is the
\emph{root mean square} (RMS) error, defined as $\frac{1}{\sqrt{mn}}\left\Vert \widehat{U}^{T}\widehat{V}-\Theta\right\Vert _{F}$.
The result is $0.17$, $0.15$ and $0.13$ respectively. This means
that the error of the estimator of the rating yielded by our algorithm
is within $\pm0.2$ in average. Note that the difference between two
successive ratings is $1$, our algorithm successfully recovers the
true ratings.

Next we move on to \emph{MovieLens} 100k dataset. This dataset is
composed of 943 users and 1682 movies. There are 100k user ratings
with the values among $\left\{ 1,2,3,4,5\right\} $. We use this dataset
to do 5-fold cross validation. During each cross validation, the whole
dataset is divided into a training set with 80\% ratings and a testing
set with 20\% ratings. This is done with the official division provided.
Our algorithm yields an error of $0.963$ slightly worse than the
best error registered $0.894$. This may be because that our model
does not take into consideration of the intercepts (i.e. preprocessing
by moving the rating average towards zero).

\bibliographystyle{unsrt}
\bibliography{DRS}

\begin{thebibliography}{10}

\bibitem{ricci2011introduction}
Francesco Ricci, Lior Rokach, and Bracha Shapira.
\newblock {\em Introduction to recommender systems handbook}.
\newblock Springer, 2011.

\bibitem{bennett2007netflix}
James Bennett and Stan Lanning.
\newblock The netflix prize.
\newblock In {\em Proceedings of KDD cup and workshop}, volume 2007, page~35,
  2007.

\bibitem{narayanan2008robust}
Arvind Narayanan and Vitaly Shmatikov.
\newblock Robust de-anonymization of large sparse datasets.
\newblock In {\em Security and Privacy, 2008. SP 2008. IEEE Symposium on},
  pages 111--125. IEEE, 2008.

\bibitem{han2004scalable}
Peng Han, Bo~Xie, Fan Yang, and Ruimin Shen.
\newblock A scalable p2p recommender system based on distributed collaborative
  filtering.
\newblock {\em Expert systems with applications}, 27(2):203--210, 2004.

\bibitem{kermarrec2010application}
Anne-Marie Kermarrec, Vincent Leroy, Afshin Moin, and Christopher Thraves.
\newblock Application of random walks to decentralized recommender systems.
\newblock In {\em Principles of Distributed Systems}, pages 48--63. Springer,
  2010.

\bibitem{massa2007trust}
Paolo Massa and Paolo Avesani.
\newblock Trust-aware recommender systems.
\newblock In {\em Proceedings of the 2007 ACM conference on Recommender
  systems}, pages 17--24. ACM, 2007.

\bibitem{o2005trust}
John O'Donovan and Barry Smyth.
\newblock Trust in recommender systems.
\newblock In {\em Proceedings of the 10th international conference on
  Intelligent user interfaces}, pages 167--174. ACM, 2005.

\bibitem{ziegler2005towards}
Cai-Nicolas Ziegler.
\newblock {\em Towards decentralized recommender systems}.
\newblock PhD thesis, Citeseer, 2005.

\bibitem{koren2009matrix}
Yehuda Koren, Robert Bell, and Chris Volinsky.
\newblock Matrix factorization techniques for recommender systems.
\newblock {\em Computer}, (8):30--37, 2009.

\bibitem{liao2010network}
Yongjun Liao, Pierre Geurts, and Guy Leduc.
\newblock Network distance prediction based on decentralized matrix
  factorization.
\newblock In {\em NETWORKING 2010}, pages 15--26. Springer, 2010.

\bibitem{liao2013dmfsgd}
Yongjun Liao, Wei Du, Pierre Geurts, and Guy Leduc.
\newblock Dmfsgd: A decentralized matrix factorization algorithm for network
  distance prediction.
\newblock {\em Networking, IEEE/ACM Transactions on}, 21(5):1511--1524, 2013.

\bibitem{recht2010guaranteed}
Benjamin Recht, Maryam Fazel, and Pablo~A Parrilo.
\newblock Guaranteed minimum-rank solutions of linear matrix equations via
  nuclear norm minimization.
\newblock {\em SIAM review}, 52(3):471--501, 2010.

\bibitem{fazel2008compressed}
M~Fazel, E~Candes, B~Recht, and P~Parrilo.
\newblock Compressed sensing and robust recovery of low rank matrices.
\newblock In {\em Signals, Systems and Computers, 2008 42nd Asilomar Conference
  on}, pages 1043--1047. IEEE, 2008.

\end{thebibliography}

\end{document}